\documentclass{journal}

\usepackage{graphics} 
\usepackage{multirow}
\usepackage{caption}
\usepackage{subcaption}
\usepackage[english]{babel}
\usepackage{epsfig} 
\usepackage{mathptmx} 
\usepackage{times} 
\usepackage{amsmath} 
\usepackage{amssymb}  
\usepackage{enumerate}
\usepackage{bm}
\usepackage{epstopdf}
\usepackage{comment}
\usepackage{bbm}
\usepackage[square, numbers, comma, sort&compress]{natbib}  
\usepackage{float}
\usepackage{algorithm}
\usepackage[noend]{algpseudocode}
\usepackage{cases}
\usepackage{algorithmicx}
\usepackage{algpseudocode}
\usepackage{placeins}
\usepackage{url}
\mdseries\itshape
\algdef{SE}[DOWHILE]{Do}{doWhile}{\algorithmicdo}[1]{\algorithmicwhile\ #1}%
\usepackage{tikz}
\usepackage{xcolor}
\usetikzlibrary{arrows, automata,positioning,calc,shapes,decorations.pathreplacing,decorations.markings,shapes.misc,petri,topaths}
  \usepackage{pgfplots}
  \pgfplotsset{compat=newest}
  \usetikzlibrary{plotmarks}
  \usepackage{grffile}
  \newlength\figureheight
  \newlength\figurewidth
\newtheorem{theorem}{Theorem}[section]

\newtheorem{lemma}[theorem]{Lemma}

\renewcommand{\P}{\mathbb P}
\newcommand{\E}{ \mathbb E}

\renewcommand{\tilde}{\widetilde}
\renewcommand{\hat}{\widehat}

\renewcommand{\rho}{\varrho}

\newtheorem{prop}{Proposition}

\newcommand{\rev}[1]{{#1}}

\definecolor{plotcolor1}{rgb}{0,0.447,0.741}
\definecolor{plotcolor2}{rgb}{0.741,0,0.447}
\definecolor{plotcolor3}{rgb}{0,0.741,0.294}
\definecolor{plotcolor4}{rgb}{0.741,0.294,0}
\definecolor{plotcoloraux}{rgb}{0.447,0.447,0.447}
\tikzset{plotstyle1/.style={color=plotcolor1,solid,line width=1.0pt}}
\tikzset{plotstyle2/.style={color=plotcolor2,densely dashed,line width=1.0pt}}
\tikzset{plotstyle3/.style={color=plotcolor3,dotted,line width=1.0pt}}
\tikzset{plotstyle4/.style={color=plotcolor4,loosely dashed,line width=1.0pt}}
\tikzset{auxlines/.style={color=plotcoloraux,solid,line width=0.5pt}}
\newcommand{\markersize}{0.7pt}
\tikzset{discretemarkers/.style={mark=*,mark options={solid},mark size=\markersize}}

\title{\LARGE \bf
The Value of Information and Efficient Switching in Channel Selection
}

\author{Jiesen Wang, Yoni Nazarathy, Thomas Taimre}
\begin{document}

\maketitle
\thispagestyle{empty}
\pagestyle{empty}

\begin{abstract}
We consider a collection of statistically identical two-state continuous time Markov chains (channels). A controller continuously selects a channel with the view of maximizing infinite horizon average reward. A switching cost is paid upon channel changes. We consider two cases: full observation (all channels observed simultaneously) and partial observation (only the current channel observed). We analyze the difference in performance between these cases for various policies. For the partial observation case with two channels, or an infinite number of channels, we explicitly characterize an optimal threshold for two sensible policies which we name ``call-gapping'' and ``cool-off''. Our results present a qualitative view on the interaction of the number of channels, the available information, and the switching costs.
\end{abstract}

\section{Introduction}

Many scenarios in mining, finance, telecommunications, medical research, and other fields, involve a situation where the reward collected from a resource fluctuates over time in a stochastic manner. For example the yield obtained from mineral resource extraction varies as digging persists; the returns from financial investments vary based on many random market effects; the communication bit rate of communication channels varies based on physical conditions; and the findings in medical trials vary over time. Such a theme of randomly varying rewards, or a randomly modulated reward rate, reappears in multiple application areas. Hence designing, managing, and controlling such uncertain situations has been a focal point of stochastic operations research in a variety of contexts. The theory of restless bandits presents one general paradigm for dealing with such problems.  \rev{See for example \citet{GGW11}, \citet{whittle1988restless}, and \citet{WW90} for the general restless bandits problem.} The main theme in such research is the efficient selection of channels/projects/arms over time.

In this paper we add to the body of literature by considering problems with switching costs. Switching costs have been considered in \citet{AHT90}, \citet{BS94}, and \citet{DH03}
in the context of multi-armed bandit processes, but to the best of our knowledge have not been studied as we do here.  The general problem which we discuss is one in which a resource yields random rewards over time, with some known average reward rate and a maximal reward rate. One way to improve performance is to introduce additional independent instances of this resource and to consider a situation where at any time we use \rev{a resource} of our (dynamic) choice. By doing so, we are potentially able to increase the obtained reward rate from the average reward rate towards the maximal reward rate. 

\rev{
As alluded to above, dynamic resource allocation problems span a variety of applications. However, one key area in which decisions often need to be made at the millisecond or second timescale is communication networks. Specifically dynamic resource allocation problems dealing with opportunistic scheduling as in \citet{ALT19a} and further in \cite{ALT19} have recently attracted attention. In such a setting wireless cellular systems utilize random channel quality
variations to 
minimize the flow-level holding costs. Some of this research has utilized the restless bandits formulation, as in \citet{JV12} (and further in \cite{ALT19a, ALT19}). Related is the continuous time paradigm with impulsive control as in \citet{AGV21}.  Switching costs often need to be taken into account and this is the key contribution of our current paper since the aforementioned works did not take switching costs directly into account.

Other related work in the context of communication networks is in the domain of handover problems. See for example the survey \citet{WXAC16} as well as \citet{MGPRZ16} where general problems dealing with handover and their solution via Markov Decision Processes (MDPs) are formulated. A key aspect of this domain is the potential lack of information with regards to channel quality. This introduces interesting problems and has attracted much work for which a survey describing contributions up to 2015 is in \citet{kuhn2015wireless}. In more recent years, several additional notable works in which channel information is not fully available are \citet{KMMM19}, \citet{LAV16} and further in \citet{LADP17}. See also recent work dealing with minimizing the age of information as \citet{MKAE20} and \citet{MK21}.
}

\rev{As an abstract example of a communication link, assume a link that yields} an average bit rate of $4$ Mbit/sec and a maximal bit rate of $10$ Mbit/sec where the actual instantaneous bit rate varies stochastically over time and achieves the maximal bit rate for random finite durations. By introducing multiple instances of such a link and allowing the system to switch between the instances without constraint, we are able to get an effective average bit rate that exceeds $4$ Mbit/sec and potentially nears the maximal bit rate of $10$ Mbit/sec. The key is clearly to use the right instance of the channel at ``the right time'', so as to on-average use channels that do better than the average bit rate. At the extreme positive case, by introducing an increasing number of instances and assuming their stochastic behavior is independent between instances, we can get arbitrarily close to the $10$ Mbit/sec bit rate. The other extreme is not switching between channels at all and settling for the average bit rate of $4$ Mbit/sec, obtained from a single channel.

\begin{figure}[h]
	\centering
	\includegraphics[scale=0.4]{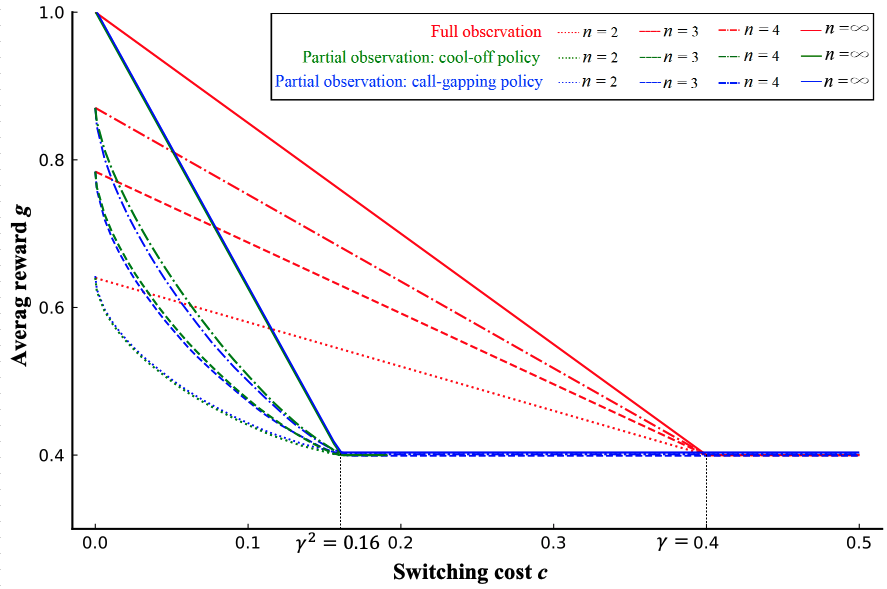}
	\caption{The optimal average reward for a $\gamma = 0.4$ system with a varying number of channels $n$. The red curves are for a case of full observation. The green curves are for a case of partial observation with a cool-off policy. The blue curves are for the case of partial observation with a call-gapping policy. Note that for $n=2$ and $n = \infty$ both of the latter policies are identical.\label{fig:2425}}
\end{figure}

While the usage of such redundant channels can be of benefit, it is clearly not without cost. First there are the structural costs of setting up additional channels. The exact nature of these costs depends on the application and is not our focus. Then there are costs associated with setting up systems for gaining information about the instantaneous state of all channels. Finally there are dynamic instantaneous switching costs involved when switching between channels. In this paper we study the tradeoffs and costs associated with this problem using the simplest example model that we could consider. The elegance of our simple model is that it captures the value of real-time information and at the same time allows us to compare how the addition of more channels to the system increases rewards. Real systems are bound to be more complex than the model which we present, however the results from our model capture the essential tradeoffs that one can expect.

After reparameterization, our model is that each of $n$ channels yield instantaneous rewards of either $0$ or $1$ where switching between the reward states is according to a two state continuous time Markov chain. The long term average reward obtained by a channel is $\gamma \in (0,1)$ and the instantaneous cost of switching between channels is $c$. We distinguish between a case of {\em full observation} where the state of all channels is observable, and a case of {\em partial observation} where only the state of the current channel is available. For each of these cases, we suggest parameterized channel selection policies, and we are able to analyze and optimize the parameters of these policies in certain situations. 

An illustration of the performance of such a model for the case of $\gamma = 0.4$ is in Figure~\ref{fig:2425}. There are two threshold values at $\gamma=0.4$ and at $\gamma^2 = 0.16$. We see that in the full observation case, when $c > \gamma$ it is not worthwhile to use additional channels and otherwise the total rewards increases in an affine manner as $c \to 0$. Further we see that in the partial observation case, usage of redundant channels is only worthwhile if $c < \gamma^2$. In this case, the total reward rate increases in a non linear manner as $c \to 0$. In this case, we compare two policies that are described in the sequel, namely {\em call-gapping} and {\em cool-off}. While we don't have optimality proofs for these policies within the class of all policies, we show that these policies agree in performance for $n=2$ and $n \to \infty$. We further obtain explicit results for the case of $n=2$ and at $n=\infty$. Note that in all cases, as the number of channels increases, the achievable reward rate increases towards the maximal reward for small switching costs $c$. Our results in this paper help understand the tradeoffs between the number of channels, switching costs, information, and policy qualitatively.

\rev{
The remainder of the paper is structured as follows: In Section~\ref{sec:model} we present the model and summarize the main results. In Section~\ref{sec:marbCaseI} we further study the full observation case by considering simple steady state results, the Whittle Index, and the optimality. In Section~\ref{sec:partialObs} we derive results associated with the case of partial observations. In Section~\ref{sec:numerical} we present numerical results, and finally we conclude in Section~\ref{sec:conclusion}.
}
\section{Model and Summary of Results}
\label{sec:model}

We consider $n$ statistically identical channels $i=1,2,\ldots,n$, with state $X_i(t) \in \{0,1\}$, evolving independently according to a continuous-time, time-homogenous, two-state Markov chain model with generator
\[
Q = \left[
\begin{array}{cc}
-\lambda  & \lambda \\
\mu & -\mu
\end{array}
\right]\,,
\]
where $\lambda, \mu >0$.

At any given time, the controller may only use one of the channels. The controller's choice is indicated through $U(t) \in \{1,\ldots,n\}$, where $U(t)=i$ means that channel $i$ is being used at time $t$. There is a fixed switching cost of $\kappa\geq 0$ for every instant at which $U(t)$ is changed; that is, every time $t$ where $U(t^-) \neq U(t)$, an additional cost of $\kappa$ is incurred.

When $X_i(t) = x \in \{0,1\}$, the reward rate of that channel is denoted by $r(x)$, a function assumed identical for all channels. The goal of the controller is then to maximize the average reward,
\[
g =\liminf_{t \to \infty} \frac{ \E \Big[ \int_0^t   r \big( X_{U(s)}(s) \big) \, ds \Big]  - \kappa \,  N(t) }{t} ,
\]
where $N(t)$ denotes the number of channel changes during the time interval $[0,t]$.

By observing that
\[
\int_0^t   r \big( X_{U(t)}(t) \big) \, dt = t \, r(0) + \big(r(1) - r(0)\big) \int_0^t X_{U(s)}(s) \, ds\,,
\]
w.l.o.g., we are able to set $r(0)=0$ and $r(1)=\zeta>0$.   Further, by rescaling time by a factor of $\eta = \lambda^{-1}$ and setting $\zeta/\eta$ to be $1$, we parameterize the channel model only by $\gamma = \lambda/(\lambda+\mu)$ (the stationary probability of a channel being in state $1$) and $c = \kappa/\eta$.

In what follows, we write $I(t) = X_{U(t)}(t)$, which denotes the state of the selected channel at time $t$.
Then, the average reward $g$ corresponding to the parametrization just outlined is expressed as:
\[
g =
\liminf_{t \to \infty}   \frac{ (\zeta/\eta) \int_0^t \tilde{I}(u) \, du - (\kappa/\eta) \tilde{N}(t)}{t}
=
\liminf_{t \to \infty}   \frac{  \int_0^t \tilde{I}(u) \, du - c \tilde{N}(t)}{t}\,,
\]
where $\tilde{I}(t) = I(\eta t)$ and $\tilde{N}(t) = N(\eta t)$. 

Hence our model parameters are $\gamma \in (0,1)$ and $c \ge 0$ with a rescaled generator
\[
Q = \left[
\begin{array}{cc}
-1  & 1\\
1/\gamma-1 & -(1/\gamma-1)
\end{array}
\right]\,.
\]
We distinguish two situations:
\begin{description}
\item {\bf Full Observation (Case I)}: The controller fully observes $\{X_1(t),\ldots,X_n(t)\}$.
\item {\bf Partial Observations (Case II)}: When $U(t) = i$ the controller only observes $X_i(t)$.
\end{description}

In considering this partial observation case, it is useful to consider belief states for $i=1,\ldots,n$, via
\[
\omega_i(t) = \P\big(X_i(t)=1 \,|\, \mbox{All information observed up to time}~t \big),
\]
and in constructing the belief states, it is useful to denote the {\em last time in channel} $i$ via,
\[
{\cal T}_i(t) = \sup \{ t' \le t ~:~ U(t') = i \}\,.
\]
This allows the construction of the belief state of channel $i$ via
\[
\omega_i(t) = p\big(t - {\cal T}_i(t) \, ; \, X_i({\cal T}_i(t))\big),
\]
where 
$p(t \,;\, x) := \P\big( X_i(t) = 1 \, | \,  X_i(0) =x \big)$  which can be represented explicitly as
\begin{equation}
\label{eq:2stateMove}
p(t \,;\, x)
 = \left\{
\begin{aligned}
& \gamma(1-e^{-\frac{t}{\gamma}}), & & x=0\,, \\
& \gamma+(1-\gamma)e^{-\frac{t}{\gamma}}, & & x=1\,.
\end{aligned}
\right.
\end{equation}
Observe that if $U(t) = i$ then ${\cal T}_i(t) = t$ and then $\omega_i(t) = X_i(t)$ which is either $0$ or $1$.

{\bf Control policies:} We consider several policies all of which limit channel switching to times $t$ when $X_i(t) = 0$. This means that in the partial observation case (II), the belief state at any time $t$ for $i \neq U(t)$ is strictly monotonically increasing in time according to the first expression in \eqref{eq:2stateMove}. In both case~I and case~II, one trivial policy denoted $\pi^{(s)}$ is the policy of never switching channel.  Further, in case~I, there is the policy $\pi^{(a)}$ which switches to an arbitrary channel $j \neq U(t)$ if $X_j(t) = 1$ and $X_{U(t)} = 0$. That is, this policy ensures that the current channel is in state $1$ if such a channel exists, and further this policy does not switch excessively if not needed.

In case~II, in addition to the policy $\pi^{(s)}$, we consider a {\em call-gapping policy}, $\pi^{(\tau)}$, and a {\em cool-off} policy $\pi^{(\sigma)}$.  In brief, call-gapping does not switch out of a channel prior to a call-gapping duration, parameter $\tau > 0$. Such a policy was analyzed in a different context in  \citet{LR04} where it was termed call-gapping. Further in brief, cool-off does not switch into a channel prior to a cooling-off period, parameter $\sigma > 0$. 

Like all our policies, call-gapping does not switch out of a good channel, however if the current channel $U(t)$ is in bad state, i.e.\ $X_{U(t)} = 0$ then this policy switches to the channel with highest belief state as long as the time since the last switch is not less than~$\tau$. That is, at time $t$ denote the {\em last switching time} via,
\[
{\cal T}^{\ell}(t) = \sup \{ t' \le t ~:~ U(t'^-) \neq U(t') \}\,,
\]
then the call-gapping policy switches only if $X_{U(t)} = 0$ and  $t- {\cal T}^\ell(t) \ge \tau$.  Note that in our context, since the channels are homogenous and $\omega_i(t)$ is monotonically increasing, when the call-gapping policy yields a switch then it is to a channel that was not used for the longest duration. This naturally yields {\em round-robin} behavior as w.l.o.g., we switch from channel $i$ to channel $i+1$ (modulo $n$). 

As with the call-gapping policy, the cool-off policy does not switch out of a good channel and once switching occurs we switch with a round-robin manner, i.e.\ to the channel that has been visited longest ago. However, in contrast to the call-gapping policy, we choose to switch into a channel only if it hasn't been visited for a time that is at least the cool-off parameter $\sigma$. This essentially means, that the cool-off policy considers the belief states of all channels and switches into a channel $i$ only if $U(t) = 0$ and $\omega_i(t)$ exceeds the threshold $p(\sigma~;~0)$. Note that unlike the call-gapping policy, the cool-off policy may potentially incur multiple instantaneous switches.  Also note that when $n=2$, the call-gapping and cool-off policies are identical.
Both the call-gapping and cool-off policies ensure a strict upper limit on the switching costs, $c \tilde{N}(t)/t$, as with these policies, almost surely,

\[
\limsup_{t \to \infty}~ c \frac{\tilde{N}(t)}{t} \le \frac{c}{\tau},
\qquad
\text{and}
\qquad
\limsup_{t \to \infty}~ c \frac{\tilde{N}(t)}{t} \le \frac{n c}{\sigma}.
\]

Our results for these policies are as follows. First for case~I, the performance of the system under $\pi^{(s)}$ and $\pi^{(a)}$ is tractable, for any value of $n$. This also allows us to characterize, when each of these policies is preferable as a function of the switching cost $c$ and the parameter $\gamma$. 

In contrast, for case~II, analysis is more difficult. When $n=2$, we are able to characterize the optimal call-gapping parameter $\tau^*$. For finite $n > 2$, we have not found such an analytic characterization. However, we can consider systems with large $n$. For this we construct an approximate model, that we label with $n = \infty$. Such a system has only two channels, say $i=1$ and $i=2$.  When $U(t) = 1$, then $X_2(t)$ is assumed to be in steady state. Similarly when $U(t) = 2$ then $X_1(t)$ is assumed to be in steady state. That is, the current channel $X_{U(t)}$ behaves normally, however at the moment of switching to the other channel, the state of that channel is drawn from the steady state distribution $[(1-\gamma)~~~\gamma]$, and the evolution continuous. Under a round-robin based policy such as our $\pi^{(\tau)}$ and $\pi^{(\sigma)}$, with large $n$, the $n = \infty$ model approximates the system because we can expect the next channel in the round-robin to be at approximate steady state.

The nature of our analytic results is in comparing the different policies (and their parameters) for Case~I, and Case~II, and for different values of $n$. That is, for Case~I, we determine when  $\pi^{(s)}$ is preferable to  $\pi^{(a)}$ and vice versa. We also determine the value of the long term reward $g$. Similarly for Case~II (under $n=2$ and $n=\infty$), we compare  $\pi^{(\tau)}$,  $\pi^{(\sigma)}$, and  $\pi^{(s)}$, determine the optimal parameters for $\tau$ (or $\sigma$), and obtain expressions for $g$.

\rev{The following two theorems are proved in Sections~\ref{sec:marbCaseI} and \ref{sec:partialObs} respectively.}

\rev{
\begin{theorem}
\label{thm1}
In Case~I, the optimal choice between $\pi^{(a)}$ and $\pi^{(s)}$ is given by:
\[
\pi^* =
\begin{cases}
\pi^{(a)}, & c < \gamma\,, \\
\pi^{(s)}, & c \ge \gamma\,.
\end{cases}
\]
Further, under $\pi^*$ the optimal expected average reward is given by:
\[
g^* = \begin{cases}
1- (1-\gamma )^n -c \frac{1- \gamma -(1-\gamma )^n}{\gamma } & c < \gamma\,, \\
\gamma & c \ge \gamma\,.
\end{cases}
\] 
Moreover, for $n=2$ the policy $\pi^*$ is an optimal policy.

\hfill$\square$
\end{theorem}
}

\rev{
We also conjecture that in Case~I $\pi^*$ is the optimal policy for any $n$. However, we are only able to prove this using computer algebra systems (such as Mathematica) for small $n$ by doing symbolic computations. Additionally as we show in Section~\ref{sec:marbCaseI}, $\pi^*$ also corresponds to the Whittle Index based policy.
}

A similar partial result is found in Case~II: for any $n$, if $c > \gamma^2$ then the optimal policy is $\pi^{(s)}$ when compared against $\pi^{(\tau)}$ and $\pi^{(\sigma)}$. We now have,

\begin{theorem}
	\label{thm2}
	For Case II, we consider two scenarios.  Firstly, restrict the policy space to be the set of call-gapping policies and denote the optimal policy therein by $\pi^*_C$.  Secondly, restrict the policy space to be the set of all cool-off policies and denote the optimal policy therein by $\pi^*_D$.  Then we have for $n \geq 2$:
	\[
	\pi^*_C =
	\begin{cases}
	\pi^{(\tau^*)}, & c < \gamma^2, \\
	\pi^{(s)}, & c \ge \gamma^2,
	\end{cases}\quad 
	\pi^*_D =
	\begin{cases}
	\pi^{(\sigma^*)}, & c < \gamma^2, \\
	\pi^{(s)}, & c \ge \gamma^2.
	\end{cases}
	\]
	When $c < \gamma^2$ and $n = 2$, the {\em optimal call-gapping time}, $\tau^*$ is the unique non-negative solution of the equation
	\begin{equation}
	\label{eqn:optTrans}
	e^{\frac{2\tau^*}{\gamma}}(\gamma^2-c)(\gamma-2)+2e^{\frac{\tau^*}{\gamma}}\gamma(\gamma-\tau^*(\gamma-1))-\gamma(\gamma^2-c) = 0.
	\end{equation}
	The optimal cool-off level is given by $\sigma^*=\tau^*$.
	When $c < \gamma^2$, $n = \infty$,
	\[
   \sigma^*= \tau^* =  0 \,.
	\]
	Further with $\pi^*_C$ or $\pi^*_D$, the optimal expected average reward is given by:
	\[
	g_C^* = g_D^* = \begin{cases}
	\frac{  A_1(\gamma,\tau^*) - c ~A_2(\gamma,\tau^*)     }{A_3(\gamma,\tau^*)}  & c < \gamma^2, n = 2 \,, \\
	1-c \, (1-\gamma)/\gamma^2 & c < \gamma^2, n = \infty \,, \\
	\gamma & c \geq \gamma^2 \,,
	\end{cases} 
	\]
	where,
	\begin{align*}
	A_1(\gamma,\tau) & = e^{\frac{2\tau}{\gamma}}((\tau-1)\gamma^3-(3\tau-2)\gamma^2+2\tau\gamma)-2e^{\frac{\tau}{\gamma}}\gamma^2(1-\gamma)^2+2\gamma^4+(\tau-3)\gamma^3-\tau\gamma^2,\\
	A_2(\gamma,\tau) & = (\gamma-1)(e^{\frac{2\tau}{\gamma}}(\gamma-2)+\gamma), \\
	A_3(\gamma,\tau) & = \gamma^3+(\tau-2)\gamma^2-\tau\gamma -e^{\frac{2\tau}{\gamma}}(\gamma^3-(\tau+2)\gamma^2+3\tau\gamma-2\tau).
	\end{align*}
	\hfill$\square$
\end{theorem}

Note that the expression for $g^*_C$ or $g^*_D$ as in Theorem~\ref{thm2} can be used to evaluate the performance for $n=2$ for any value of $\tau$ (alt. $\sigma$), by replacing $\tau^*$ in the expression with $\tau$ (alt. $\sigma$). Also note that empirical evidence, see Figure~\ref{fig:2425}, suggests that,  when $2 < n <\infty$ that the optimal cool-off policy has better expected  average reward than the optimal call-gapping policy.

\rev{
\section{On the Whittle Index and Optimality for Case~I}
\label{sec:marbCaseI}

We now prove Theorem~\ref{thm1} in two parts; first the expression for $g^*$ is derived by using a continuous time birth and death process. Then we prove that $\pi^*$ is the optimal policy for $n=2$, describe symbolic computations for proving it is optimal for higher order $n$, and conjecture it is optimal for any $n$. We then move onto the Whittle index associated with Case~I.

}

{\bf Proof of correctness of $g^*$ expression for Theorem~\ref{thm1}:}
We consider the process $\{M(t), \, t \ge 0\}$ on the state space $\{0,1,\ldots,n\}$ where,
\[
M(t) = \sum_{i=1}^n X_i(t).
\]
The fact that the channels are independent and satisfy the same probability law, implies that $M(t)$ is a birth and death continuous time Markov chain, with birth and death rates 
\begin{equation}
\label{eq:bdRates}
\lambda_i = n-i,
\qquad
\mu_i = i \Big(\frac{1}{\gamma} - 1 \Big),
\end{equation}
for $i \in  \{0,\ldots,n-1\}$ and $i \in \{1,\ldots,n\}$ respectively. Such a birth and death process is known as an Ehrenfest model with binomial stationary distribution,
\rev{
\[
\rho_i =   {{n}\choose{i}} \gamma^i \,(1-\gamma)^{n-i} \qquad i = 0, \ldots, n.
\]
}

Under the policy $\pi^{(a)}$ a reward rate of $1$ is accrued at times $t$ during which $M(t) > 0$. Further, switching costs are incurred at certain transition times of $M(t)$. Specifically at times $t$ when $M(t^-) = 0$ and $M(t) = 1$ a switching cost $c$ is incurred with probability $(n-1)/n$. This is the probability that the channel that changes to `good' is not the currently selected channel and therefore an immediate channel change takes place. Similarly, at times $t$ when $M(t^-) = k \ge 2$ and $M(t) = k-1$, a cost of $c$ is incurred with probability $1/k$. This is the probability of the current channel turning `bad' and hence requiring a switch. Now considering the reward as a Markov reward process, we have that average under policy $\pi^{(a)}$ is,
\rev{
\begin{align*}
g &=  (1-\rho_0) - c \frac{n-1}{n}\lambda_0 \rho_0 - c \sum_{k=2}^n \frac{1}{k} \mu_k \rho_k \\
&=1- (1-\gamma )^n -c \frac{1- \gamma -(1-\gamma )^n}{\gamma },
\end{align*}
}
where the second line follows from the structure of the birth and death rates \eqref{eq:bdRates}. Further with $\pi^{(s)}$ we have that $g$ is trivially $\gamma$. Hence $\pi^{(a)}$ is preferable only when $c \le \gamma$.
\hfill $\square$

\vspace{20pt}

\rev{
To consider the optimality of $\pi^*$, the processes $\{X_i(t)\}_{i=1}^n$ and $U(t)$ can be transformed to $\tilde{W}(t)$ and $\tilde{S}(t)$ where $\tilde{W}(t)= X_{U(t)}(t)$ is the state of the current channel and,
\[
\tilde{S}(t) = \sum_{i \neq U(t)} X_i(t),
\]
is the number of other channels that are in the good state. The process
\[
\tilde{{\cal X}}(t) = \big(\tilde{W}(t), \tilde{S}(t) \big),
\]
can then be represented as part of a continuous time MDP with singular controls, similar to the framework used in \citet{AGV21}. The state space is $\{0,1\} \times \{0,\ldots,n-1\}$. The action process $\{\tilde{A}(t)\}$ takes values in the action space $\{0,1\}$ where $\tilde{A}(t) = 0$ indicates staying on a channel at time $t$. Further at singular time points, $t$ in which $\tilde{A}(t)=1$, an instantaneous channel switch is made (and immediately at $t^+$, $\tilde{A}(t^+) = 0$) . In the construction of this process, we assume that if a switch is made and there is another channel available of a different state to $\tilde{W}(t)$, then we switch into such an arbitrary channel and this implies state change. Otherwise (in the border cases in which a switch is made and there isn't a different channel), the state is not changed.

To construct the continuous time MDP with singular controls, we use $I^{(a)}(w,s)$ as an indicator if transitions out of state $(\tilde{W}(t),\tilde{S}(t))=(w,s)$ are singular or not for action $a \in \{0,1\}$. If $I^{(a)}(w,s) = 0$ then we define transition intensities, $\tilde{q}^{(a)}((w,s),\cdot)$. Alternatively if  $I^{(a)}(w,s) = 1$ then there are transition probabilities, $\tilde{p}^{(a)}((w,s),\cdot)$. In our case, it is simply that $I^{(a)}(w,s) = a$ and hence for action $a=0$ (don't switch) we have transition intensities and for action $a=1$ (switch) we have transition probabilities. These are constructed via,

\[
\begin{array}{rclcr}
\tilde{q}^{(0)}( (w,s), (w,s+1) ) &= & \tilde{\lambda}_s   &  \text{for} & s \in \{0,\ldots,n-2\},  \\ 
\tilde{q}^{(0)}( (w,s), (w,s-1) ) & = & \tilde{\mu}_s   &\text{for}&s \in \{1,\ldots,n-1\}, \\ 
\tilde{q}^{(0)}( (0,s), (1,s) )  & = & 1 &   \text{for} & s \in \{0,\ldots,n-1\}, \\ 
\tilde{q}^{(0)}( (1,s), (0,s) ) & = & \frac{1}{\gamma} - 1 &   \text{for} & s \in \{0,\ldots,n-1\}, \\ 
\end{array}
\]
where
\[
\tilde{\lambda}_s = n-1-s,
\qquad
\tilde{\mu}_s = s \Big(\frac{1}{\gamma} - 1 \Big),
\]
and $\tilde{q}^{(0)}((w,s),\cdot) = 0$ otherwise. Further,
\begin{align*}
\tilde{p}^{(1)}( (0,s), (1,s-1) ) & = 1  \qquad \text{for} \qquad s \in \{1,\ldots,n-1\}, \\ 
\tilde{p}^{(1)}( (1,s), (0,s+1) ) & = 1 \qquad \text{for} \qquad s \in \{0,\ldots,n-2\}, \\ 
\tilde{p}^{(1)}( (0,0), (0,0) ) & = 1 , \\ 
\tilde{p}^{(1)}( (1,n-1), (1,n-1) ) & = 1, 
\end{align*}
and $\tilde{p}^{(1)}((w,s),\cdot) = 0$ otherwise. 

This continuous time MDP with singular controls can be converted to a continuous time MDP without singular controls using the same approach as in \citet{AGV21}. Further we can uniformize with a uniformization rate of $n/\gamma$ to obtain a standard discrete time MDP with the same state space and action space. In this MDP we denote the state process via
\[
{{\cal X}}(k) = \big({W}(k), {S}(k) \big),
\qquad
k=0,1,2,\ldots
\]
and the action sequence denoted $\{A(n)\}$ takes values in $\{0,1\}$. The transition probabilities for this MDP are,
\[
\begin{array}{rclcl}
p^{(0)}( (w,s), (w,s+1) ) &= & \lambda_s   &  \text{for} & s \in \{0,\ldots,n-2\},  \\ [5pt]
p^{(0)}( (w,s), (w,s-1) ) & = & \mu_s  &\text{for}&s \in \{1,\ldots,n-1\}, \\[5pt]
p^{(0)}( (0,s), (1,s) )  & = & \frac{\gamma}{n} &   \text{for} & s \in \{0,\ldots,n-1\}, \\ [5pt]
p^{(0)}( (1,s), (0,s) ) & = & \frac{1-\gamma}{n} &   \text{for} & s \in \{0,\ldots,n-1\}, \\ [5pt]
p^{(0)}( (w,s), (w,s) ) & = & \frac{(w+s) \gamma +(n-w-s)(1-\gamma)}{n} &   \text{for} & s \in \{0,\ldots,n-1\}, \\ [5pt]
p^{(1)}( (0,s), (1,s) ) &= & \lambda_{s-1}  &  \text{for} & s \in \{1,\ldots,n-1\},  \\ [5pt]
p^{(1)}( (0,s), (1,s-2) ) & = & \mu_{s-1}   &\text{for}&s \in \{2,\ldots,n-1\}, \\[5pt]
p^{(1)}( (0,s), (0,s-1) )  & = & \frac{1-\gamma}{n} &   \text{for} & s \in \{1,\ldots,n-1\}, \\ [5pt]
p^{(1)}( (0,s), (1,s-1) ) & = & \frac{s\gamma + (n-s)(1-\gamma)}{n} &   \text{for} & s \in \{1,\ldots,n-1\}, \\ [5pt]
p^{(1)}( (1,s), (0,s+2) ) &= & \lambda_{s+1}  &  \text{for} & s \in \{0,\ldots,n-3\},\quad (*)  \\ [5pt]
p^{(1)}( (1,s), (0,s) ) & = & \mu_{s+1}   &\text{for}&s \in \{0,\ldots,n-2\}, \\[5pt]
p^{(1)}( (1,s), (1,s+1) ) & = & \frac{\gamma}{n}   &\text{for}&s \in \{0,\ldots,n-2\}, \\[5pt]
p^{(1)}( (1,s), (0,s+1) )  & = & \frac{(s+1)\gamma+(n-s-1)(1-\gamma)}{n} &   \text{for} & s \in \{0,\ldots,n-2\}, \\ 
\end{array}
\]

where,
\[
\lambda_s = \gamma \frac{n-1-s}{n},
\qquad
\mu_s =   (1-\gamma) \frac{s}{n},
\]
and ${p}^{(a)}((w,s),\cdot) = 0$ otherwise. Note that the transition above marked via $(*)$ only occurs for systems with $n\ge3$.

The rewards of this uniformized MDP are,
\begin{align*}
r^{(0)}(w,s) &= \frac{\gamma}{n}w,\\
r^{(1)}(w,s) &=
\begin{cases}
\frac{\gamma}{n} - c & \text{if} \quad \Big( w=0 \quad \text{and} \quad s>0\Big)~~\text{or}~~\Big(w=1 \quad \text{and} \quad s=n-1 \Big),  \\
-c  & \text{otherwise}. \\
\end{cases}
\end{align*}

We are now able to prove that the optimal policy for this MDP is $\pi^*$ in the case of~$n=2$.

{\bf Proof of optimality of $\pi^*$ for $n=2$ in Theorem~\ref{thm1}:}

We carry out policy evaluation for this MDP under $\pi^*$ and this yields a relative value function, $V^*: \{0,1\} \times\{0,1\} \to {\mathbb R}$, interpreted as,

\[
V^*(w,s) = \lim_{\ell \to \infty} \E [ \Big(\sum_{k=0}^{\ell-1} r^{\big(A(k)\big)}\big(W(k),S(k) \big)\Big) - \ell \overline{g}^*_{(w,s)} ~|~ W(0) = w, \, S(0) = s],
\]
where,
\[
\overline{g}^*_{(w,s)} = \lim_{\ell \to \infty} \E [ \frac{1}{\ell} \sum_{k=0}^{\ell-1} r^{\big(A(k)\big)}\big(W(k),S(k) \big)  ~|~ W(0) = w, \, S(0) = s].
\]

Note that this MDP is a uni-chain and thus $\overline{g}^*_{(w,s)}$ does not depend on the initial state and can be denoted as $\overline{g}^*$. Now we carry out explicit policy evaluation (under the policy $\pi^*$) where we constrain $V^*(0,0) = 0$. This is done separately for $c<\gamma$ (where $\pi^* = \pi^{(a)}$) and $c \ge \gamma$ (where $\pi^* = \pi^{(s)}$). Resulting expressions for the relative value function and gains are,
\[
\begin{array}{ll}
& V^*(0,1) =  \frac{(c-\gamma)(\gamma - 2)}{2}, \\
& V^*(1,0) =  \frac{\gamma(c - \gamma + 2)}{2}, \\
& V^*(1,1)	=  \frac{3\gamma -c}{2} + c\gamma - \gamma^2, \\
& \overline{g}^* = \frac{\gamma (1-(1-\gamma)^2-c(1-\gamma))}{2},
\end{array}
\qquad
\text{for}
\qquad
c < \gamma,
\]
and
\[
\begin{array}{ll}
& V^*(0,1) =  0, \\
& V^*(1,0) =  \gamma, \\
& V^*(1,1)	=  \gamma, \\
& \overline{g}^* = \frac{\gamma^2}{2},
\end{array}
\qquad
\text{for}
\qquad
c < \gamma.
\]
Note that as expected, taking $n=2$, we have that $\overline{g}^* = \frac{\gamma}{n} g^*$, where $g^*$ is as in Theorem~\ref{thm1}.

We now consider the Bellman equation for the MDP,
\begin{equation}
\label{eq:bellman1}
V + g \, \bm{1} = \max \{\underbrace{r^{(0)} + P^{(0)} V}_{R_0},~
\underbrace{r^{(1)} + P^{(1)} V}_{R_1}
\},
\end{equation}
where $\bm{1}$ is a vector of ones, $V$ and $r^{(\cdot)}$ are the appropriate vectors and $P^{(\cdot)}$ is the appropriate transition probability matrix (using lexicographic ordering of the state space). Using $V=V^*$ from above, expressions for the vector difference $R_1-R_0$ are as follows:
\[
\begin{array}{cl}
c & \text{for} \quad(w,s) = (0,0),\\
\frac{c - \gamma}{2} & \text{for} \quad (w,s) = (0,1),\\
\frac{3c+\gamma}{2} & \text{for}\quad (w,s) = (1,0),\\
c & \text{for} \quad (w,s) = (1,1),\\
\end{array}
\qquad
\text{for}
\qquad
c < \gamma,
\]
and
\[
\begin{array}{cl}
c & \text{for} \quad(w,s) = (0,0),\\
c - \gamma & \text{for} \quad (w,s) = (0,1),\\
c+\gamma & \text{for}\quad (w,s) = (1,0),\\
c & \text{for} \quad (w,s) = (1,1),\\
\end{array}
\qquad
\text{for}
\qquad
c \ge \gamma.
\]
The respective signs of these entries are $(+,-,+,+)$ for $c< \gamma$ and $(+,+,+,+)$ for $c \ge \gamma$.  These signs agree with $\pi^*$ meaning that the only case for switching is in state $(0,1)$ when $c < \gamma$. Hence, $V^*$ is indeed an optimal solution of \eqref{eq:bellman1} because it was computed for $\pi^*$ which agrees with these signs. Hence $\pi^*$ is an optimal policy.
\hfill $\square$

\vspace{20pt}

We note that using computer algebra systems (Mathematica in our case), we are able to explicitly carry out policy evaluation and obtain expressions for $V^*$ for $n=2,3,4,\ldots,8$ in both the $c <\gamma$ and $c \ge \gamma$ cases\footnote{Higher values of $n$ can potentially be handled as well with greater computation cost (we did not carry this out).}.  In all these cases, we can also compute $R_1-R_0$ and see that the signs of the entries agree with $\pi^*$ and hence $\pi^*$ is optimal. Note that these computations also involve finding an expression for the gain, $\overline{g}^*$. In all cases we obtain $\overline{g}^* = \frac{\gamma}{n} g^*$ ($g^*$ is as in Theorem~\ref{thm1}), similarly to the case $n=2$ in the proof above .

 Hence for $n \in \{2,\ldots,8\}$ we essentially have a proof of optimality (in addition to the explicit proof written above for $n=2$). However, we were not able to distill explicit expressions for $V^*$ using a simple formula and hence we do not have a proof for arbitrary $n$. Nevertheless, we conjecture that $\pi^*$ is optimal in Case~I for any $n$. See the paper's associated GitHub repository \citet{WNT20} for the associated Mathematica code.

\subsection*{The Whittle Index}

It is also interesting to consider the system as if controlled by the Whittle index (see \citet{GGW11} for an overview). As this is a continuous time system with impulsive controls we use a setup similar to \citet{AGV21}.

In general an index based policy operates as follows. There is an index function ${\cal I}: {\chi} \to {\mathbb R}$, where ${\chi}$ is the state space of the channel. The index based policy continuously monitors the index of each channel, say ${\cal I}_i$ for $i=1,\ldots ,n$ by evaluating ${\cal I}(\cdot)$ on the channel state. It then ensures that the chosen channel is the channel $i$ with the highest index ${\cal I}_i$. In our continuous time setup, we assume that the policy only switches channels if necessary. That is, if there is a tie, there isn't a switch.

The Whittle index is one such index function. A key component for evaluation of the Whittle index is the one arm subsidy problem, parameterized via $\nu \in {\mathbb R}$. In this problem, instead of considering the full system of $n$ channels we focus on a single channel and assume the controller has a choice between two actions: $a=0$ which means not using the channel and receiving a subsidy at rate $\nu$, or $a=1$ which means using the channel. 

In the one arm subsidy problem for our case, the state space is 
\[
\chi = \big\{(x,u)~:~x \in \{0,1\}, ~u \in \{0,1\} \big\},
\]
where $x$ and $u$ are indicators of whether the channel is in a good state and whether the channel is in use, respectively.

Since we are dealing with impulsive controls, similarly to the $n$ channel continuous time impulsive MDP above, we use $I^{(a)}(x,u)$ as indicators for impulsive control. Here, in the one arm subsidy problem, $I^{(a)}(x,u) = \mathbbm{1}_{\{a \neq u\}}$. With these we have transition rates (for state-action pairs without impulsive controls),
\begin{align*}
	&\tilde{q}^{(0)}((0,0),(1,0))   = 1,  &   \tilde{q}^{(0)}( (1,0),(0,0)) = \frac{1}{\gamma}-1, \\
	&\tilde{q}^{(1)}( (0,1),(1,1)) = 1,   & \tilde{q}^{(1)}( (1,1),(0,1)) = \frac{1}{\gamma}-1, \\
\end{align*}
and $\tilde{q}^{(\cdot)}((x,u),\cdot) = 0$ otherwise for all other state-action pairs without impulsive control. Further, the transition probabilities (for state-action pairs with impulsive control) are,
\begin{align*}
	&\tilde{p}^{(0)}((0,1),(0,0)) = 1,	& \tilde{p}^{(0)}((1,1),(1,0)) = 1, \\
	& \tilde{p}^{(1)}((0,0),(0,1) ) =1,	& \tilde{p}^{(1)}((1,0),(1,1) ) = 1, 
\end{align*}
and $\tilde{p}^{(\cdot)}((x,u),\cdot) = 0$ otherwise for all other state-action pairs with impulsive control. 

In such a one-arm subsidy problem, the continuous time reward is taken as 
\begin{equation}
	\tilde{r}^{(a)}(x,u)   = x\,u \qquad x \in \{0,1\} ,~~ u \in \{0,1\}\,.
\end{equation}
Further in adapting the one-arm subsidy problem to our general control problem we set instantaneous rewards at transitions where $I^{(a)}(x,u) =1$ at a value of $-c/2$. Here the factor of $1/2$ captures the fact that a one-arm subsidy problem is not a real problem being used but rather an abstraction of a single channel vs. all channels together. In the real system, every channel switch would result in a cost of $c$ (reward of $-c$) and then the one-arm subsidy problem of the current channel ``becomes'' the problem of the next channel. 

%
%




The continuous time MDP with singular controls of the one-arm subsidy problem can now be converted to a continuous time MDP without singular controls using the same approach as \citet{AGV21} and our analysis above. Further, similar to the MDP above, we can uniformize, now using a uniformization rate of $1/\gamma$. This yields a standard discrete time MDP with the same state space and action space. The transition probabilities for this uniformized MDP have for every $(x,u) \in \chi$,
\begin{equation}
\label{eq:whittle-values-p}
\begin{array}{rcl}
p^{(0)}((x,u),(0,0)) &=& 1-\gamma,\\
p^{(0)}((x,u),(1,0)) &=& \gamma,
\end{array}
\qquad
\text{and}
\qquad
\begin{array}{rcl}
p^{(1)}((x,u),(0,1)) &=& 1-\gamma,\\
p^{(1)}((x,u),(1,1)) &=& \gamma,
\end{array}
\end{equation}
and $p^{(\cdot)}((x,u),\cdot) = 0$ otherwise.
Further, the rewards are now,
\[
r^{(a)}(x,u) =  
\gamma \, x  - {\mathbf 1}_{\{a \neq u\}}\frac{c}{2}.
\]
Observe that the first term captures the uniformized continuous time reward and the second term captures the impulsive costs.

Now for any fixed $\nu \in {\mathbb R}$, consider the Bellman equation for relative value function $V^\nu(x,u)$,

\begin{equation}
\label{eq:bellman-whittle}
V^\nu + g^\nu \, \bm{1} = \max \{
\underbrace{r^{(0)} + \nu \, \bm{1} + P^{(0)} V^\nu}_{R_0^\nu},~
\underbrace{r^{(1)} + P^{(1)} V^\nu}_{R_1^\nu}
\},
\end{equation}
where as previously, $\bm{1}$ is a vector of ones, $V^\nu$ and $r^{(\cdot)}$ are the appropriate $4$-vectors and $P^{(\cdot)}$ is the appropriate transition probability matrix (using lexicographic ordering of the state space) using \eqref{eq:whittle-values-p}, and $g^\nu$ is the (scalar) long term average reward for the problem parameterized by $\nu$.

It can now be verified that the Whittle index for $c < \gamma\,$ is:
\begin{equation}
\label{eq:whittle-solutions-A}
\nu(0,0) = 0, \qquad 
\nu(0,1) = c \, \gamma, \qquad 
\nu(1,0) = c \, \gamma + (\gamma-c), \qquad \nu(1,1) = \gamma\,.
\end{equation}
And further the Whittle index for $c \ge \gamma\,$ is:
\begin{equation}
\label{eq:whittle-solutions-B}
\nu(0,0) = 0, \qquad 
\nu(0,1) = \gamma^2 \,, \qquad 
\nu(1,0) = \gamma^2, \qquad \nu(1,1) = \gamma\,.
\end{equation}

That is, for each state $(x,u) \in \chi$, $\nu(x,u)$ in \eqref{eq:whittle-solutions-A} or \eqref{eq:whittle-solutions-B}  defines a Bellman equation \eqref{eq:bellman-whittle} which has a solution in which $R_0^\nu(x,u) = R_1^\nu(x,u)$ (again using lexicographic ordering for the vectors $R_0^\nu$ and $R_1^\nu$). This can be verified, also by noting that the associated value function solutions $(V^{\nu},g^\nu)$ of \eqref{eq:bellman-whittle} are as follows for $c < \gamma\,$:
\[
\begin{array}{ll}
\Big(\begin{bmatrix}
	0 & \frac{c}{2} & \gamma & \gamma + \frac{c}{2} 
\end{bmatrix}^\top 
,
\gamma^2\Big)
& \text{for} \quad\nu(0,0) = 0, \\[+6pt]
\Big(\begin{bmatrix}
0 & -\frac{c}{2} & \gamma-c & \gamma -\frac{c}{2} 
\end{bmatrix}^\top
,
\gamma^2\Big) & \text{for} \quad\nu(0,1) = c \, \gamma, \\[+6pt]
\Big(\begin{bmatrix}
0 & -\frac{c}{2} & 0 &  \frac{c}{2} 
\end{bmatrix}^\top
,
\gamma - c + c\,\gamma\Big) & \text{for} \quad\nu(1,0) = c \, \gamma + (\gamma-c) , \\[+6pt]
\Big(\begin{bmatrix}
0 & -\frac{c}{2} & 0 &  -\frac{c}{2} 
\end{bmatrix}^\top
,
\gamma\Big) & \text{for} \quad \nu(1,1) = \gamma\,. \\[+6pt]
\end{array} 
\]

And when $c \ge \gamma\,$ these are,

\[
\begin{array}{ll}
\Big(\begin{bmatrix}
	0 & \frac{c}{2} & \gamma & \gamma + \frac{c}{2} 
\end{bmatrix}^\top 
,
\gamma^2\Big)
& \text{for} \quad\nu(0,0) = 0, \\[+6pt]
\Big(\begin{bmatrix}
0 & -\frac{c}{2} & 0 & \gamma -\frac{c}{2}
\end{bmatrix}^\top
,
\gamma^2\Big) & \text{for} \quad\nu(0,1) =  \gamma^2, \\[+6pt]
\Big(\begin{bmatrix}
0 & \frac{c}{2} -\gamma& 0 &  \frac{c}{2} 
\end{bmatrix}^\top
,
\gamma^2\Big) & \text{for} \quad\nu(1,0) = \gamma^2 , \\[+6pt]
\Big(\begin{bmatrix}
0 & -\frac{c}{2} & 0 &  -\frac{c}{2} 
\end{bmatrix}^\top
,
\gamma\Big) & \text{for} \quad \nu(1,1) = \gamma\,. \\[+6pt]
\end{array} 
\]


With these Whittle index expressions at hand, we now observe that a Whittle index based policy agrees with $\pi^*$ of Theorem~\ref{thm1}.

\begin{prop}
Consider Case~I, and assume the system is controlled via a Whittle index policy, which ensures that at any time $t$ the chosen channel is $U(t)$ with the highest index as given by \eqref{eq:whittle-solutions-A} for $c < \gamma$ or \eqref{eq:whittle-solutions-B} for $c \ge \gamma$,  and performs switching only when necessary. Then the system operating under this policy is equivalent to a system operating under $\pi^*$ of Theorem~\ref{thm1}.
\end{prop}

\noindent
{\bf Proof:}
Observe that both in the $c < \gamma$ and $c \ge \gamma$ cases,
\[
\nu(1,0) < \nu(1,1),
\qquad
\text{and}
\qquad
\nu(0,0) < \nu(1,1).
\] 
Hence the system controlled by a Whittle index policy never switches out of a good channel $(X_{U(t)}(t) = 1)$. 

Further, at time $t$ if $X_{U(t^-)}(t^-) = 1$ and then $X_{U(t)}(t) = 0$ (the state of the current channel turns from the good state to the bad state), then if for all the other channels $j \neq U(t)$, $X_j(t) = 0$, then the index based policy chooses to stay with $U(t)$, since both in the $c < \gamma$ and $c \ge \gamma$ cases, $\nu(0,0) < \nu(0,1) $. 

Conversely in such a situation, if there is at least one good channel in the system, say $i \neq U(t)$ with $X_i(t) = 1$, then the index based policy chooses to switch to channel $i$ (or any other good channel) only if $c < \gamma$, and otherwise stays. This is because, $\nu(0,1) < \nu(1,0)$ only if  $c < \gamma$ and $\nu(0,1) = \nu(1,0)$ when $c \ge \gamma$.
\hfill $\square$

}
\section{Renewal Reward Analysis  for Case II}
\label{sec:partialObs}

In this section we prove Theorem~\ref{thm2} via a regenerative analysis for $n=2$ and later for $n=\infty$. We consider the regenerative structure based on regeneration points where switching has just occurred into a bad channel ($=0$). That is, take the time-axis and consider time points at which the controller switched (from a bad channel) into a bad channel. These are regeneration points which we denote by $T_0=0, T_1, T_2,\ldots$. We can assume that at time $t=0$ the system starts at such a point because we are looking at the infinite horizon average cost. Note that between $T_{k-1}$ and $T_k$ it is possible that there were other switches --- namely switches into a channel that is in good state ($=1$).

The net reward obtained during the interval $(T_{k-1},\,T_k]$, is given by 
\[
V_k = \int_{T_{k-1}}^{T_k} \tilde{I}(u) \, du - c \big(\tilde{N}(T_k) - \tilde{N}(T_{k-1})\big).
\]
That is the total time during the interval {when} the channel was in good state less the switching costs.

From the regenerative property of the process, it turns out that the sequence $\big(T_{k}-T_{k-1},\, V_k \big)$ is an i.i.d.\ sequence. We denote $W$ as a generic random variable distributed as $T_{k}-T_{k-1}$ and $V$ as a generic random variable distributed as $V_k$. It then follows from the ``Renewal Reward Theorem'', see for example Theorem 1.2, Chap VI \citet{A08}, that the average reward is given by
\begin{equation}
\label{eq:g}
g = \frac{\E[V]}{\E[W]}\,.
\end{equation}
Note that by construction $W$ is non-lattice and both $V$ and $W$ have a finite mean.
Hence our goal is now to compute the expectation of $V$ and $W$ under a call gaping policy with parameter $\tau$. For $n=2$ or $n=\infty$ this is equivalent to the cool-off policy with $\sigma = \tau$.

We now construct two generic random variables via their probability distributions that come to aid. We denote these as $W_0$ and $W_1$ and their CDFs by $F_{W_i}(t)$ for $i=0,1$. We have,
\begin{equation}
\label{eq:Fwi}
F_{W_i}(t) = \mathbb{P}(W_i \le t) =  
\begin{cases}
0,  & t < \tau, \\
(1-p(\tau \,;\, i)) + p(\tau \,;\, i) (1- e^{-\frac{1-\gamma}{\gamma}(t-\tau)}), & \tau \le t.
\end{cases}
\end{equation}
These random variables are mixtures of a mass at $\tau$ and a shifted exponential. The random variable $W_i$ denotes the time of switching under a policy with call-gapping parameter $\tau$ when at time $0$ a switch occurred into a channel in state $i$. It is constructed by observing that if at time $t=\tau$ the state is $0$ we switch with probability $1-p(\tau \,;\, i)$; otherwise we wait an exponentially distributed duration until switching. Note that,
\begin{equation}
\label{eq:expectWi}
\E[W_i] = \tau + p(\tau \,;\, i) \frac{\gamma}{1-\gamma}.
\end{equation}

With the generic random variables $W_0$ and $W_1$ at hand, we can analyse a complete regenerative cycle of duration $W$. For this denote,
\[
W=\tilde{W}_0+\tilde{W}_1+\tilde{W}_2+\ldots + \tilde{W}_M,
\]
with $\tilde{W}_i$ denoting the inter-switching times within the cycle. Here $M$ is a random variable with support $0,1,2,\ldots$, denoting the number of times we switch into a good channel within such a cycle. It then holds by construction that,
\[
\tilde{W}_0~ =^d ~W_0,
\qquad \qquad
\tilde{W}_i~ =^d~ W_1,~~ i=1,2,\ldots.
\]
This follows since our regeneration points are the time points at which the controller switched from a bad channel into a bad channel. Then each cycle starts with a duration distributed as $W_0$. It is then followed by $M$ additional durations which are switches from bad states to good states, each distributed as $W_1$. 

The following two lemmas yield explicit expressions for the denominator and numerator of \eqref{eq:g}.

\begin{lemma}
The denominator of \eqref{eq:g} can be represented as,
\begin{align*}
\E[W] &= \E[W_0]+\frac{\E[p(W_0 \,;\, 0)]}{1-\E[p(W_1 \,;\, 0)]}\E[W_1]\\
&=\frac{e^{\frac{2\tau}{\gamma}}(\gamma^3 - (\tau+2) \gamma^2 +3\tau \gamma-2\tau) - \gamma^3-(\tau-2) \gamma^2+\tau\gamma}{(\gamma-1)^2(e^{\frac{2\tau}{\gamma}} (\gamma-2)-2 e^{\frac{\tau}{\gamma}}\gamma+\gamma)}.
\end{align*}
 \hfill$\square$
\end{lemma}
\noindent{\bf Proof:}

Observe that,
\[
W =^d \tilde{W}_0+\tilde{I}_0 \tilde{W}_1+\tilde{I}_0\tilde{I_1} \tilde{W}_2+\cdots+\tilde{I}_0 \tilde{I}_1\cdots \tilde{I}_{k-1}\tilde{W}_k+\cdots\,,
\]
where the indicator, $\tilde{I}_i$, for $i=0,1,2,\ldots$ is equal to $1$ if the jump at the end of the interval associated with $\tilde{W}_i$ was to a good channel. Note that the random variables $\tilde{I}_i$ for $i=1,2,\ldots$ are identically distributed and $\tilde{I}_0$ follows a different distribution. We can now denote generic random variables, $I_0$ and $I_1$ satisfying
\[
\tilde{I}_0~ =^d ~I_0\,,
\qquad \qquad
\tilde{I}_i~ =^d~ I_1,~~ i=1,2,\ldots\,.
\]
For these two random variables, by conditioning on $W_i$ (for $i=0,1$), it holds that
\begin{align}
\notag
\E[I_i] &= \E[ \E[I_i \, | \, W_i] ] \\
\notag
	&=  \E[p(W_i \, ; \, 0)] \\
\notag
&= (1-p(\tau \, ;\, i))p(\tau \, ;\, 0)+p(\tau \, ;\, i)\int_{\tau}^{\infty}p(t \, ;\, 0) \frac{1-\gamma}{\gamma} e^{-\frac{1-\gamma}{\gamma} (t-\tau)} \, dt\\
\label{eq:expectIi}
&=\gamma-\frac{e^{-\frac{\tau}{\gamma}}\gamma(\gamma+p(\tau \, ;\, i)-2)}{\gamma-2},
\end{align}
where the second step follows because every switch is out of a bad channel and the third step follows from  \eqref{eq:Fwi}.

Observe that the sequence $\{\tilde{I}_i\}$ is a mutually independent sequence of random variables and for each $i$, $\tilde{I}_i$ is independent of $\tilde{W}_j$ for all $j \neq i$. Based on the fact that $\E[\tilde{W}_i]$ is same as $\E[W_1]$ for $i=1,2,\ldots$, we have
\begin{equation}
\label{eq:geomStuff}
\begin{aligned}
\E[W]&=\E[W_0]+\E[I_0]\E[W_1]+\E[I_0]\E[I_1]\E[W_1]+ \ldots + \E[I_0](\E[I_1])^{k-1}\E[W_1]+\cdots\\
 &=\E[W_0]+\E[I_0]\frac{1}{1-\E[I_1]}\E[W_1]\,.
\end{aligned}
\end{equation}
After manipulation using \eqref{eq:2stateMove}, \eqref{eq:expectWi}, and \eqref{eq:expectIi}, the result follows.
\hfill$\square$
 
\begin{lemma}
The numerator, $\E[V]$ of \eqref{eq:g} can be represented as
\[
\E[V] = \E[R] - c \, (\E[M]+1),
\]
where,
\begin{align*}
\E[R]&=\E[R_0]+\frac{\E[p(W_0 \,;\, 0)]}{1-\E[p(W_1 \,;\, 0)]}\E[R_1] \\
&= \frac{e^{\frac{2\tau}{\gamma}} ((1-\tau)\gamma^3+(3\tau-2)\gamma^2-2\tau\gamma)+2 e^{\frac{\tau}{\gamma}}\gamma^2(\gamma-1)^2-2\gamma^4+(3-\tau)\gamma^3+\tau\gamma^2}{(\gamma-1)^2(e^{\frac{2\tau}{\gamma}} (\gamma-2)-2 e^{\frac{\tau}{\gamma}}\gamma+\gamma)},
\end{align*}
and,
\begin{align*}
\E[M]&=\frac{\E[p(W_0 \,;\, 0)]}{1-\E[p(W_1 \,;\, 0)]}, \\
&=\frac{e^{\frac{2\tau}{\gamma}} (2-\gamma)-\gamma}{(e^{\frac{2\tau}{\gamma}} (\gamma-2)-2 e^{\frac{\tau}{\gamma}}\gamma+\gamma)(\gamma-1)}-1.
\end{align*}
 \hfill$\square$
\end{lemma}

\noindent{\bf Proof:}
The proof follows similar lines to the previous proof. We have that $V = R - c \, (M+1)$  with $M$ as defined above and the reward $R$ being the net time during $[0,W]$ in which a good channel is selected. As with the analysis of the cycle duration, $W$, we denote,
\[
R = \tilde{R}_0 + \tilde{R}_1 + \tilde{R}_2 + \ldots +\tilde{R}_M,
\]
where $\tilde{R}_i$ is the reward accumulated over the period corresponding to $\tilde{W}_i$. Similar to the previous analysis, we construct generic random variables $R_0$ and $R_1$ with
\[
\tilde{R}_0~ =^d ~R_0,
\qquad \qquad
\tilde{R}_i~ =^d~ R_1, i=1,2,\ldots.
\]
The expectations of these can be computed to be
\begin{align*}
\E[R_0]&=\int_{0}^{\tau}p(t \, ; \, 0)dt+p(\tau \, ; \, 0)\frac{1-\gamma}{\gamma}=\gamma(\tau-p(\tau \, ; \, 0))+\frac{p(\tau \, ; \, 0)(1-\gamma)}{\gamma},\\
\E[R_1]&=\int_{0}^{\tau}p(t \, ; \, 1)dt+p(\tau \, ; \, 1)\frac{1-\gamma}{\gamma} =\gamma(\tau-p(\tau \, ; \, 1)+1)+\frac{p(\tau \, ; \, 1)(1-\gamma)}{\gamma}.
\end{align*}

Using $\tilde{I}_i$ as in the previous proof, we observe,
\[
R =^d \tilde{R}_0+\tilde{I}_0 \tilde{R}_1+\tilde{I}_0\tilde{I_1} \tilde{R}_2+\cdots+\tilde{I}_0 \tilde{I}_1\cdots \tilde{I}_{k-1}\tilde{R}_k+\cdots\,,
\]
where again for any $i$, $\tilde{I}_i$ is independent of $\tilde{R}_j$ for $j \neq i$. Now a similar geometric series to \eqref{eq:geomStuff} is applied. The expression for $M$ is computed in a similar manner by observing
\[
M =^d \tilde{I}_0 + \tilde{I}_0\tilde{I_1} + \tilde{I}_0\tilde{I_1}\tilde{I}_2 + \ldots.
\]  
 \hfill$\square$
 
\noindent
We can now prove Theorem~\ref{thm2} for $n=2$.\\
{\bf Proof:}
Combining and manipulating the explicit expressions from the lemmas above we obtain that under a policy $\pi^{(\tau^*)}$ (alt. $\pi^{(\sigma^*)}$), the reward as in \eqref{eq:g} as a function of $\tau$ (alt. $\sigma$) is,

\begin{equation}
\label{eq:gTau}
g(\tau)  = \frac{  A_1(\gamma,\tau) - c ~A_2(\gamma,\tau)     }{A_3(\gamma,\tau)}.
\end{equation}
where $A_i(\cdot,\cdot)$, $i=1,2,3$ are as defined in Theorem~\ref{thm2}. We consider first the case $c \ge \gamma^2$. Observe that at $c = \gamma^2$,
\[
g(\tau)\Big|_{c = \gamma^2} = \gamma - \frac{2e^{\frac{\tau}{\gamma}}(\gamma-1)^2\gamma^2}{A_3(\gamma,\tau)} < \gamma
\]
and hence the $\pi^{(s)}$ policy is preferable to the $\pi^{(\tau)}$ policy for any finite $\tau$ (note that as $\tau \to \infty$ the inequality above becomes an equality). Further $g(\tau)$ is monotonically decreasing in $c$ and hence for $c > \gamma^2$ it remains optimal to use $\pi^{(s)}$.

Moving to the $c < \gamma^2$ case, we optimize the (continuous and differentiable) function $g(\tau)$ over $(0,\infty)$ to obtain equation \eqref{eqn:optTrans} from Theorem~\ref{thm2}. We do this by representing the derivative as $g'(\tau) = h(\tau) f(\tau)$ where,
\[
h(\tau) =  \frac{(\gamma-1)^2(e^{\frac{2\tau}{\gamma}}(2-\gamma)+\gamma)}{(\gamma((\gamma-2)\gamma+\tau(1-\gamma))+e^{\frac{2\tau}{\gamma}}(2-\gamma)(\gamma^2+\tau-\tau\gamma))^2},
\]
and,
\[
f(\tau) =  e^{\frac{2\tau}{\gamma}}(\gamma^2-c)(\gamma-2)+2e^{\frac{\tau}{\gamma}}\gamma(\gamma-\tau(\gamma-1))-\gamma(\gamma^2-c).
\]
It can be shown that for $c < \gamma^2$ the derivative $f'(\tau) < 0$.  Now since $h(\cdot)>0$ we have that, $g'(\tau) = 0$ if and only if $f(\tau) = 0$ and equation~\eqref{eqn:optTrans} is $f(\tau^*) = 0$. Since $f(0)> 0$ and $\lim_{\tau \to \infty}f(\tau)<0$, it is evident that for $c<\gamma^2$ there is a single root to $f(\tau) = 0$ and further at the solution $\tau^*$ the second order conditions hold,

\[
g^{''}(\tau^*)= 2h({\tau^*})\Big((1-e^{\frac{{\tau^*}}{\gamma}})\gamma^2 -c - e^{\frac{{\tau^*}}{\gamma}} {\tau^*}(1-\gamma) \Big) < 0.
\]

\hfill $\square$

Note that at the extremes of $\tau$, the reward $g(\tau)$ as in \eqref{eq:gTau} yields the expected results. With zero switching costs,
\[
\lim_{\tau \to 0} g(\tau)\Big|_{c = 0} = \lim_{\tau \to 0}\frac{A_1(\gamma,\tau)}{A_3(\gamma,\tau)} = 1 - (1-\gamma)^2,
\]
and further without switching,
\[
\lim_{\tau \to \infty} \frac{A_1(\gamma,\tau)-c A_2(\gamma,\tau)}{A_3(\gamma,\tau)} = \lim_{\tau \to \infty} \frac{A_1(\gamma,\tau)}{A_3(\gamma,\tau)}-c\,0=\gamma.
\]

In the case of $n=\infty$ a similar (yet simpler) analysis to all of the above section pursues. As described in Section~\ref{sec:model} the system may be viewed as a two channel system where every transition is always to a steady state channel. The usage of the transient probabilities $p(\tau~;~i)$ as in \eqref{eq:Fwi} is then replaced by $\gamma$. This results in correspondently simpler expressions in all the cases where transient probabilities are used ($n=2$) and can now be replaced by $\gamma$. The resulting reward expression is then,
\[
g(\tau) = \frac{(\gamma -1)c+\gamma (\gamma -\gamma  \tau+\tau)}{\gamma ^2-\gamma  \tau+\tau},
\]
with the derivative, 
\[
g'(\tau) = \frac{(\gamma -1)^2 \left(c-\gamma ^2 \right)}{\left(\gamma ^2-\gamma  \tau+\tau\right)^2}.
\]
This shows that $g(\cdot)$ is monotonic decreasing in $\tau$ if $c <\gamma^2$ and monotonic increasing in $\tau$ if $c > \gamma^2$. Hence for $c < \gamma^2$ the optimal switching parameter is $\tau^* = 0$ and yields reward as in Theorem~\ref{thm2}. Further, since $\lim\limits_{\tau \rightarrow \infty}g(\tau) = \gamma$, for $c > \gamma^2$ it is not optimal to use $\pi^{(\tau)}$ for any finite $\tau$, and $\pi^{(s)}$ is preferable. 

\section{Numerical Results for Case II}
\label{sec:numerical}

We now carry out numerical experiments for our system under various policies, focusing on Case~II. We first describe the numerical computations and simulation experiment used for Figure~\ref{fig:2425} and then expand with further numerical results. 

The Case~I curves in the figure are obtained directly from Theorem~\ref{thm1}. The Case~II curves for $n=\infty$ are created directly using Theorem~\ref{thm2} and the $n=2$ curves are created by numerically solving equation \eqref{eqn:optTrans} of Theorem~\ref{thm2} (using the bisection method). The remaining curves for $n=3$ and $n=4$ using both the call-gapping and the cool-off policies are obtained via Monte-Carlo simulation and optimization to find the optimal $\tau$ and $\sigma$ parameters for these policies. 

This Monte-Carlo simulation (as well as other code) can be found in the GitHub repository \citet{WNT20}. We simulated the system over a grid of $c$ in the range $[0,\gamma^2]$ where $\gamma^2 = 0.16$, with a spacing of $0.005$. Then for each value of $c$ we simulated both the call-gapping and cool-off policies over respective grids of $\tau$ and $\sigma$ in the range $[0,1.55]$, each with a spacing of $0.001$. In each of these individual simulation runs we kept a constant seed to reduce variability in the curves using common random numbers, and simulated for a time horizon of $T=10^5$, with an arbitrary fixed initial condition. We then obtained the average reward $g$ using a crude Monte Carlo estimator over the continuous time $[0,T]$ and then plotted the reward for the optimal $\tau$ and $\sigma$ in each case.

In addition to the simulations leading to Figure~\ref{fig:2425} we also carried out extensive additional numerical experiments for more parameter values. Key results from these experiments are summarized in Table~\ref{tab:table1}. Our purpose is to compare the cool-off and call-gapping policies for finite $n > 2$. As we observe from Figure~\ref{fig:2425}, the cool-off policy is able to achieve slightly higher average reward for $\gamma = 0.4$. We now considered the system for several additional $\gamma$ values and several values of $n$. 

  \begin{table}[h!]
	\begin{center}
		\caption{Evaluating the difference between the cool-off and call-gapping policy.}
		\label{tab:table1}
		\begin{tabular}{|c|c|l|l|} 
			\hline
			\textbf{System} & \textbf{Number of channels} & \textbf{Maximal Gap}& \textbf{Worst cost} \\
			\hline
			\multirow{18}{*}{ $\gamma = 0.2 $} 
			&  $n=3 $ &  $0.00385 \pm 0.00029 $ &  $c = 0.0048 $\\ 
			&  $n=4 $ &  $0.00823 \pm 0.00033 $ &  $c = 0.0048 $ \\
			&  $n=5 $ &  $0.01295 \pm 0.00045 $ &  $c = 0.0056 $ \\
			&  $n=6 $ &  $0.01767 \pm 0.00045 $ &  $c = 0.0056 $ \\
			&  $n=7 $ &  $0.02188 \pm 0.00044 $ &  $c = 0.0056 $ \\
			&  $n=8 $ &  ${ 0.02619} \pm 0.00047 $ &  $c = 0.0064 $ \\
			&  $n=9 $ &  ${  0.02879} \pm 0.00051 $ &  $c = 0.0072 $ \\
			&  $n=10 $ &  ${ 0.03136} \pm 0.00057 $ &  $c = 0.0072 $ \\
			&  $n=11 $ &  ${ 0.03277} \pm 0.00047 $ &  $c = 0.0072 $ \\
			&  $n=12 $ &  ${ 0.03414} \pm 0.0006 $ &  $c = 0.008 $ \\
			&  $n=13 $ &  ${ 0.03554} \pm 0.00048 $ &  $c = 0.008 $ \\
			&   ${\bf n=14} $ &  ${\bf 0.03609 \pm 0.0006} $ &  $c = 0.0088 $ \\
			&  $ {\bf n=15} $ &  ${\bf 0.03611 \pm 0.00072} $ &  $c = 0.0096 $ \\
			&  $ {\bf n=16} $ &  ${\bf 0.03611 \pm 0.00061} $ &  $c = 0.0104 $ \\
			&  $n=17 $ &  ${ 0.03532} \pm 0.00064 $ &  $c = 0.0112 $ \\
			&  $n=18 $ &  ${ 0.0339} \pm 0.00066 $ &  $c = 0.0112 $ \\
			&  $n=19 $ &  ${ 0.0332} \pm 0.00064 $ &  $c = 0.0112 $ \\
			&  $n=20 $ &  ${ 0.03208} \pm 0.0007 $ &  $c = 0.0128 $ \\
			\hline
			\multirow{13}{*}{ $\gamma = 0.3$} 
			&  $n=3 $ &  $0.00679 \pm 0.00035 $ &  $c = 0.009 $\\ 
			&  $n=4 $ &  $0.01318 \pm 0.00044 $ &  $c = 0.009 $ \\
			&  $n=5 $ &  ${ 0.01927} \pm 0.00063 $ &  $c = 0.018 $ \\
			&  $n=6 $ &  $0.02351 \pm 0.00055 $ &  $c = 0.0144 $ \\
			&  $n=7 $ &  $0.0268 \pm 0.00074 $ &  $c = 0.018 $ \\
			&  $n=8 $ &  ${ 0.02829} \pm 0.00073 $ &  $c = 0.0198 $ \\
			&  ${\bf n=9} $ &  ${\bf 0.03012 \pm 0.00072} $ &  $c = 0.0216 $ \\
			&  $n=10 $ &  $0.02924 \pm 0.00056 $ &  $c = 0.0216 $ \\
			&  $n=11 $ &  $0.0293 \pm 0.00084 $ &  $c = 0.0252 $ \\
			&  $n=12 $ &  $0.02798 \pm 0.00096 $ &  $c = 0.0288 $ \\
			&  $n=13 $ &  $0.026 \pm 0.00078 $ &  $c = 0.0288 $ \\
			&  $n=14 $ &  $0.02524 \pm 0.00078 $ &  $c = 0.0306 $ \\
			&  $n=15 $ &  $0.02216 \pm 0.00074 $ &  $c = 0.027 $ \\
			\hline
			\multirow{8}{*}{ $\gamma = 0.4$} 
			&  $n=3 $ &  $0.00899 \pm 0.0005 $ &  $c = 0.0224 $\\ 
			&  $n=4 $ &  $0.01606 \pm 0.00066 $ &  $c = 0.032 $ \\
			&  $n=5 $ &  ${ 0.02114} \pm 0.00064 $ &  $c = 0.0256 $ \\
			&  $n=6 $ &  ${ 0.02289} \pm 0.00072 $ &  $c = 0.032 $ \\
			&  ${\bf n=7} $ &  ${\bf 0.02426 \pm 0.00071} $ &  $c = 0.0416 $ \\
			&  $n=8 $ &  $0.02294 \pm 0.0007 $ &  $c = 0.0352 $ \\
			&  $n=9 $ &  $0.0213 \pm 0.00074 $ &  $c = 0.0448 $ \\
			&  $n=10 $ &  $0.01849 \pm 0.00076 $ &  $c = 0.0544 $ \\
			
			\hline
			\multirow{3}{*}{ $\gamma = 0.6 $} 
			&  $n=3 $ &  $0.01001 \pm 0.00053 $ &  $c = 0.0648 $\\ 
			&  ${\bf n=4} $ &  ${ \bf 0.01274 \pm 0.00055} $ &  $c = 0.0648 $ \\
			&  $n=5 $ &  $0.01267 \pm 0.00064 $ &  $c = 0.1008 $ \\
			\hline
			\multirow{3}{*}{ $\gamma = 0.8 $} 
			&  ${\bf n=3 }$ &  ${\bf 0.00452 \pm 0.00042} $ &  $c = 0.1792 $\\ 
			&  $n=4 $ &  $0.00389 \pm 0.0005 $ &  $c = 0.2432 $ \\
			&  $n=5 $ &  $ { 0.00236} \pm 0.00035 $ &  $c = 0.2048 $ \\
			\hline
		\end{tabular}
	\end{center}
\end{table}
\FloatBarrier

In each case, the cool-off policy is indeed slightly better than call-gapping. The numerical results in Table~\ref{tab:table1} report the maximal gap, as well as error estimates, over a range of $c$ values between the rewards of the policies. We also present an estimate of the $c$ value in which the gap is maximal and highlight in bold the values of $n$ for which the gap is largest\footnote{For $\gamma = 0.2$ the maximal gap is at either $n=14$, $n=15$, or $n=16$ and in considering the error estimates, the exact value is not determined based on the simulations we ran.}. As an overarching summary, we see that while the cool-off policy is slightly better, the difference between the rewards is never high and hence in practice using the simpler call-gapping policy is probably preferable.  As expected from Figure~\ref{fig:2425}, for each value of $\gamma$ there is some finite $n$ in which the maximal gap is highest (for $n=2$ and $n \to \infty$ there is no gap as the policies are identical). It is also evident that for low $\gamma$ values, the worst case $n$ is higher than that of higher $\gamma$ values. This observation informed our selection of the range of $n$ for which to run the experiments. 

The computational experiments required non-negligible compute time because for any system setting $(\gamma,n,c)$ optimization over the best $\tau$ (resp. $\sigma$) parameter was required. We thus relied on an empirical observation associated with the nature of the system. We observed that for any fixed $\gamma$ and $c$, as $n$ increases, the optimal parameter ($\tau^*$ or $\sigma^*$) decreases with $n$ (with the optimal value tending to $0$ as $n \to \infty$). This allowed us to use Theorem~\ref{thm2} to first compute the optimal parameter for $n=2$ and use it as an upper bound for the (stochastic) search for the optimal parameter for $n=3$. Further for any $n=k \ge 4$, we used the estimated optimal parameter of $n=k-1$ to determine an upper bound. This allowed to use a fixed size search grid of size $20$ for the optimal $\tau$ (resp. $\sigma$) of each level $n \ge 3$. Other elements of the simulation involved the search grid over $c$. For this, after initial experimentation indicating the location of the maximal gap, we always considered $c \in [0,\gamma^2/2]$ and searched over $25$ points within this grid. Each sample path involved a time horizon of $10^3$ time units, and for each parameter setting $(\gamma,n,c)$, we simulated $100$ repetitions, allowing us to obtain $95\%$ error bounds using a standard normal approximation. The simulation duration was in the order of $24$ hours on a standard laptop using the (compiled) Julia language. 

\begin{figure}[h!]
	\centering
	\includegraphics[scale=0.55, trim={0cm 7cm 0 3cm}]{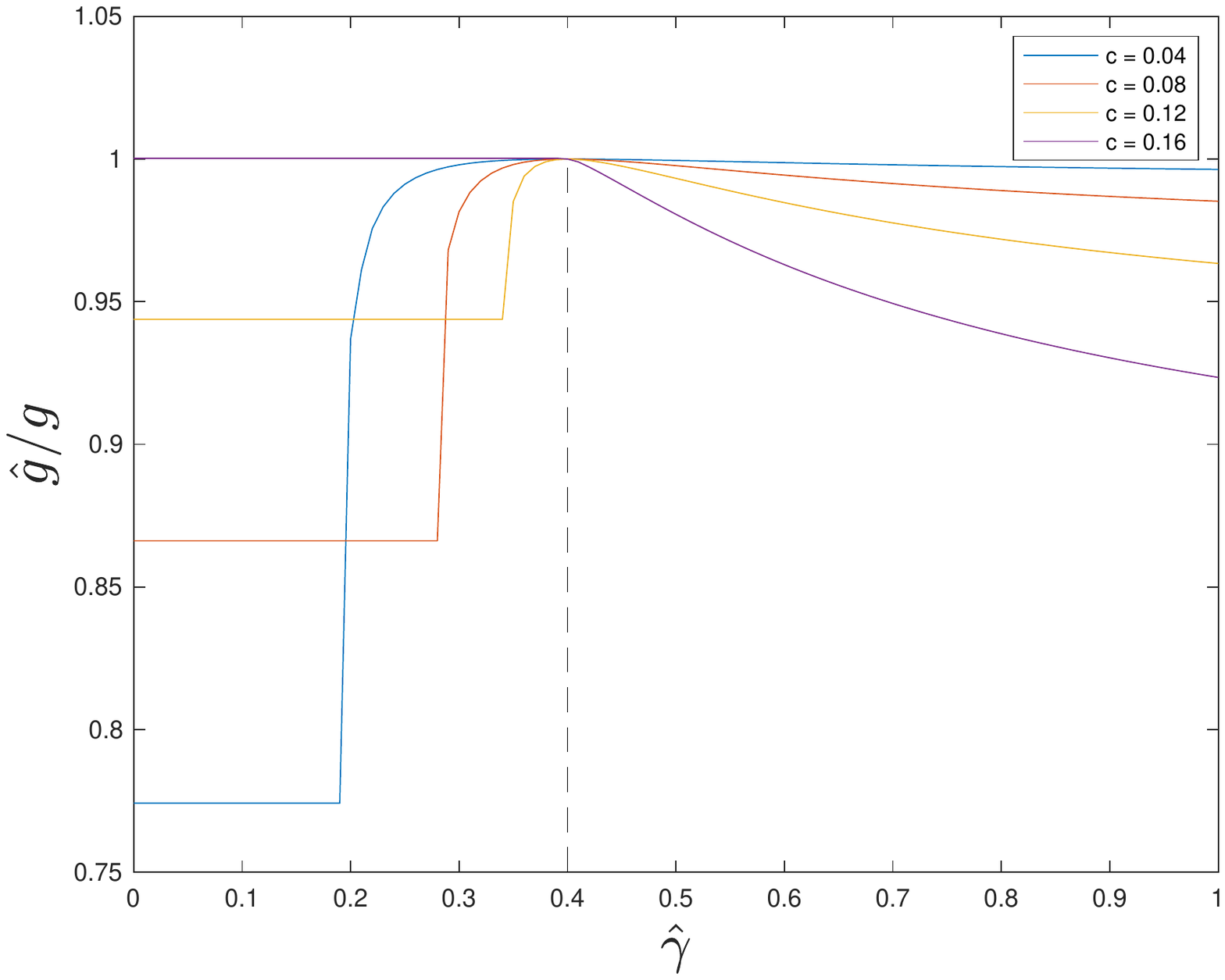}
	\caption{Robustness analysis when $\gamma = 0.4$ for Case~II and $n=2$ under call-gapping. The system is controlled with a perceived value of $\hat{\gamma}$ and the ratio of the reward and the optimal reward is plotted for several values of $c$. \label{fig:robust}}
\end{figure}

In addition to the simulations, we also carried out a robustness analysis for Case~II with $n=2$. In this case, the optimal call-gapping (and cool-off) parameter $\tau$ is easily obtained via equation \eqref{eqn:optTrans}. However we also wish to see how misspecification of the system parameter $\gamma$, leading to a misspecification of the optimal $\tau$, affects performance. For this we again consider the case where $\gamma = 0.4$, however we now assume that the system is perceived to operate at a potentially different system parameter, namely $\hat{\gamma}$. The controller then optimizes $\tau$ based on \eqref{eqn:optTrans} and the perceived value, $\hat{\gamma}$. We then compare the ratio of the actual reward, $\hat{g}$ and the ideal reward, ${g}$ obtained using $\gamma = 0.4$. Both are obtained via the expressions in Theorem~\ref{thm2}. The ratio is plotted in Figure~\ref{fig:robust} where we considered cost values $c = 0.04, 0.08, 0.12$, and $0.16$. 
As we see in Figure~\ref{fig:robust}, as we might expect, for larger deviations of $\hat{\gamma}$ from $\gamma = 0.4$ we get larger relative losses of the reward. However, the relative loss at a few percentage points in most cases. The plateaus observed on each of the curves are due to values of $\hat{\gamma}$ during which $c > \hat{\gamma}^2$ in which case the $\pi^{(s)}$ policy is employed.

\section{Conclusion and Extensions}
\label{sec:conclusion}

Our aim with this study was to obtain a qualitative view on the interaction of the number of channels, the available information, and the switching costs in a channel selection context. In such a setting, system designers can consider the value of information as well as the effect of switching costs (the value of efficient switching) in lieu of various control policies. Through a simple model, we obtained a qualitative relationship between the various system factors and in certain cases we were able to explicitly analyze the system. 

The case of full observation is straightforward to analyze, however for the case of partial observation, to the best of our knowledge, explicit analysis is only attainable with $n=2$ and $n=\infty$ as we have done here. This is for the call-gapping and cool-off policies that we introduced. Further numerical analysis was carried out for other small finite $n$ illustrating that in practice there are only minor efficiency gains to be had by using cool-off instead of call-gapping. This is despite the fact that call-gapping policy only requires local information of the current state of the channel while the cool-off policy requires information for all channels. We complement our analysis with a numerical robustness experiment hinting that inexact knowledge of system parameters would not hinder performance greatly when using a call-gapping policy. 

Our work did not focus on optimality of the policies discussed within the class of all possible policies. We conjecture that for $n=2$ call-gapping (equivalent to cool-off) is optimal in the case of partial observation. However proving this remains a challenge for a future study. Further, real systems will typically exhibit a more complicated structure than two-state Markov chains. For such systems, adaptations of the call-gapping and cool-off policies may still be employed, however analysis is more complicated. Nevertheless, we believe that our qualitative and quantitative results based on two state Markov chains may help serve as a guide for the tradeoffs between information, switching costs, policy complexity, and the number of channels in a system. 

\vspace{19pt}

\noindent
{\bf Acknowledgment:} Jiesen Wang would like to thank the University of Melbourne for supporting her work through the Melbourne Research Scholarship and the Albert Shimmins Fund. Thomas Taimre is supported by the Australian Research Council (ARC) Centre of Excellence for the Mathematical and Statistical Frontiers (ACEMS). Yoni Nazarathy and Thomas Taimre are supported by ARC grant DP180101602.


\end{document}